\documentclass[%
 reprint,
superscriptaddress,
amsmath,amssymb,
aps,
noeprint,
]{revtex4-2}
\usepackage{color}
\usepackage[utf8]{inputenc}
\usepackage{graphicx}
\usepackage{dcolumn}
\usepackage{bm}
\usepackage{upgreek}

\newcommand{\ezero}{\mbox{$E_0$}}
\newcommand{\mtwonue}{\mbox{$m^2_\nu$}}
\newcommand{\mnue}{\mbox{$m_\nu $}}

\newcommand{\mnui}{\mbox{$m_{i}$}}
\newcommand{\mtwohad}{\mbox{$|M^2_{\rm nucl}|$}}
\newcommand{\me}{\mbox{$m_\mathrm{e}$}}
\newcommand{\bspec}{$\upbeta$-spectrum}
\newcommand{\belec}{$\upbeta$-electron}
\newcommand{\bdecay}{$\upbeta$-decay}
\newcommand{\krm}{$^\mathrm{83m}$Kr}

\newcommand{\rhods}{$\rho d \cdot \sigma$}
\newcommand{\eloss}{$\varepsilon(\delta E)$}
\newcommand{\rbg}{$R_\mathrm{bg}$}

\newcommand{\ssyst}{$\sigma_{\mathrm{syst}}$}
\newcommand{\sstat}{$\sigma_{\mathrm{stat}}$}

\newcommand{\rbeta}{$R_{\upbeta}$(E)}
\begin{document}

\preprint{APS/123-QED}

\title{An improved upper limit on the 
neutrino mass \\ from a direct kinematic method by KATRIN}


\newcommand{\bonn}{Helmholtz-Institut f\"{u}r Strahlen- und Kernphysik, Rheinische Friedrich-Wilhelms-Universit\"{a}t Bonn, Nussallee 14-16, 53115 Bonn, Germany}
\newcommand{\fulda}{University of Applied Sciences~(HFD)~Fulda, Leipziger Str.~123, 36037 Fulda, Germany}
\newcommand{\etp}{Institute of Experimental Particle Physics~(ETP), Karlsruhe Institute of Technology~(KIT), Wolfgang-Gaede-Str. 1, 76131 Karlsruhe, Germany}
\newcommand{\ikp}{Institute for Nuclear Physics~(IKP), Karlsruhe Institute of Technology~(KIT), Hermann-von-Helmholtz-Platz 1, 76344 Eggenstein-Leopoldshafen, Germany}
\newcommand{\ipe}{Institute for Data Processing and Electronics~(IPE), Karlsruhe Institute of Technology~(KIT), Hermann-von-Helmholtz-Platz 1, 76344 Eggenstein-Leopoldshafen, Germany}
\newcommand{\itep}{Institute for Technical Physics~(ITEP), Karlsruhe Institute of Technology~(KIT), Hermann-von-Helmholtz-Platz 1, 76344 Eggenstein-Leopoldshafen, Germany}
\newcommand{\ppq}{Project, Process, and Quality Management~(PPQ), Karlsruhe Institute of Technology~(KIT), Hermann-von-Helmholtz-Platz 1, 76344 Eggenstein-Leopoldshafen, Germany    }
\newcommand{\inr}{Institute for Nuclear Research of Russian Academy of Sciences, 60th October Anniversary Prospect 7a, 117312 Moscow, Russia}
\newcommand{\lbnl}{Institute for Nuclear and Particle Astrophysics and Nuclear Science Division, Lawrence Berkeley National Laboratory, Berkeley, CA 94720, USA}
\newcommand{\madrid}{Departamento de Qu\'{i}mica F\'{i}sica Aplicada, Universidad Autonoma de Madrid, Campus de Cantoblanco, 28049 Madrid, Spain}
\newcommand{\mainz}{Institut f\"{u}r Physik, Johannes-Gutenberg-Universit\"{a}t Mainz, 55099 Mainz, Germany}
\newcommand{\massit}{Laboratory for Nuclear Science, Massachusetts Institute of Technology, 77 Massachusetts Ave, Cambridge, MA 02139, USA}
\newcommand{\mpik}{Max-Planck-Institut f\"{u}r Kernphysik, Saupfercheckweg 1, 69117 Heidelberg, Germany}
\newcommand{\muenster}{Institut f\"{u}r Kernphysik, Westf\"alische Wilhelms-Universit\"{a}t M\"{u}nster, Wilhelm-Klemm-Str. 9, 48149 M\"{u}nster, Germany}
\newcommand{\npi}{Nuclear Physics Institute of the CAS, v. v. i., CZ-250 68 \v{R}e\v{z}, Czech Republic}
\newcommand{\unc}{Department of Physics and Astronomy, University of North Carolina, Chapel Hill, NC 27599, USA}
\newcommand{\washington}{Center for Experimental Nuclear Physics and Astrophysics, and Dept.~of Physics, University of Washington, Seattle, WA 98195, USA}
\newcommand{\wuppertal}{Department of Physics, Faculty of Mathematics and Natural Sciences, University of Wuppertal, Gau{\ss}str. 20, 42119 Wuppertal, Germany}
\newcommand{\saclay}{IRFU (DPhP \& APC), CEA, Universit\'{e} Paris-Saclay, 91191 Gif-sur-Yvette, France }
\newcommand{\mpp}{Max-Planck-Institut f\"{u}r Physik, F\"{o}hringer Ring 6, 80805 M\"{u}nchen, Germany}
\newcommand{\cmu}{Department of Physics, Carnegie Mellon University, Pittsburgh, PA 15213, USA}
\newcommand{\cwru}{Department of Physics, Case Western Reserve University, Cleveland, OH 44106, USA}
\newcommand{\tunl}{Triangle Universities Nuclear Laboratory, Durham, NC 27708, USA}
\newcommand{\tum}{Technische Universit\"{a}t M\"{u}nchen, James-Franck-Str. 1, 85748 Garching, Germany}
\newcommand{\ornl}{Also affiliated with Oak Ridge National Laboratory, Oak Ridge, TN 37831, USA}
\newcommand{\berlin}{Institut f\"{u}r Physik, Humboldt-Universit\"{a}t zu Berlin, Newtonstr. 15, 12489 Berlin, Germany}
%

\affiliation{\ikp}
\affiliation{\itep}
\affiliation{\tum}
\affiliation{\saclay}
\affiliation{\bonn}
\affiliation{\etp}
\affiliation{\massit}
\affiliation{\muenster}
\affiliation{\ipe}
\affiliation{\mpp}
\affiliation{\mpik}
\affiliation{\mainz}
\affiliation{\unc}
\affiliation{\tunl}
\affiliation{\wuppertal}
\affiliation{\washington}
\affiliation{\npi}
\affiliation{\cmu}
\affiliation{\lbnl}
\affiliation{\fulda}
\affiliation{\inr}
\affiliation{\cwru}
\affiliation{\madrid}
\affiliation{\ppq}


\author{M.~Aker}\affiliation{\ikp}\affiliation{\itep}
\author{K.~Altenm\"{u}ller}\affiliation{\mpp}\affiliation{\tum}\affiliation{\saclay}
\author{M. Arenz}\affiliation{\bonn}
\author{M.~Babutzka}\affiliation{\etp}
\author{J.~Barrett}\affiliation{\massit}
\author{S.~Bauer}\affiliation{\muenster}
\author{M.~Beck}\affiliation{\muenster}\affiliation{\mainz}
\author{A.~Beglarian}\affiliation{\ipe}
\author{J.~Behrens}\affiliation{\etp}\affiliation{\ikp}\affiliation{\muenster}
\author{T.~Bergmann}\affiliation{\mpp}\affiliation{\tum}\affiliation{\ipe}
% ...

\collaboration{KATRIN Collaboration}

\date{\today}

\begin{abstract} 
We report on the neutrino mass measurement result from the first four-week science run of the Karlsruhe Tritium Neutrino experiment KATRIN in spring 2019. Beta-decay electrons from a high-purity gaseous molecular tritium source are energy analyzed by a high-resolution MAC-E filter. A fit of the integrated electron spectrum over a narrow interval around the kinematic endpoint at 18.57 keV gives an effective neutrino mass square value of $(-1.0~ ^{+~ 0.9}_ {-~ 1.1})$~eV$^2$. From this we derive an upper limit of 1.1~eV (90$\%$ confidence level) on the absolute mass scale of neutrinos. This value coincides with the KATRIN sensitivity.
It improves upon previous mass limits from kinematic measurements by almost a factor of two and provides 
model-independent input to cosmological studies of structure formation.

\end{abstract}

\maketitle

{\em Introduction}.-- The observation of flavor oscillations of atmospheric and solar neutrinos \cite{Fukuda:1998mi,Ahmad:2002jz} as well as oscillation studies at reactors and accelerators  unequivocally prove neutrinos to possess non-zero rest masses (e.g. \cite{Esteban:2018azc}), contradicting the Standard Model (SM) expectation of them being massless. The absolute values \mnui\ of the neutrino mass states $\nu_i$ ($i = 1,2,3$), which cannot be probed by oscillations, are of fundamental importance in cosmological studies \cite{Hannestad:2006zg,Lesgourgues:2014zoa,Loureiro:2018pdz} and for particle physics models beyond the SM \cite{King2014}. 

Due to the unique role of primordial neutrinos in the formation of large-scale structures in the universe, observations of matter clustering in different epochs of the universe allow one to probe the neutrino mass sum $\Sigma_i \mnui $. The current upper limits depend on the selection of data sets included in the analyses and are valid only within the $\Lambda$CDM concordance model \cite{Aghanim:2018eyx,Loureiro:2018pdz}.
Another model-dependent method is provided by the search for neutrinoless double beta-decay $0\nu\upbeta\upbeta$, a process forbidden in the SM due to lepton number violation. It gives access to the effective Majorana neutrino mass (e.g. \cite{Barabash:2018rds,doi:10.1146/annurev-nucl-101918-023407}).

A model-independent, direct method to probe the neutrino mass scale in the laboratory is provided by kinematic studies of weak-interaction processes such as \bdecay\ of tritium ($^3$H) and electron capture on holmium ($^{163}$Ho)   \cite{Otten2008,Drexlin2013,Gastaldo:2017edk,Nucciotti:2015rsl,Esfahani:2017dmu}. These investigations yield an incoherent sum of spectra, containing the squares of the neutrino eigenmasses $m^2_i$ as parameters. Each spectral component is weighted by the absolute square of the corresponding electron-flavor matrix element $|U_{\mathrm{e}i}|^2$. 
In the quasi-degenerate regime $m_i > 0.2$~eV, the eigenmasses are the same to better than 3~\%. The mass measured in \bdecay\ or electron capture, often called ``$m(\nu_\mathrm{e})$'', is the neutrino mass $\mnue \approx m_i$ in this regime.

Due to its low endpoint energy (\ezero\ = 18.57 keV) and favorable half-life (t$_{1/2}$ = 12.32~yr), the decay of tritium $^3$H $\rightarrow$ 
$^3$He$^+$ + e$^-$ + $\bar{\nu}_e$ has been investigated by a large number of experiments looking for the small, characteristic shape distortion of the \bspec\ close to \ezero\ due to \mnue\ \cite{Otten2008, Drexlin2013}. Experimental advances over many decades have steadily increased the sensitivity to the present upper limit of $\mnue < 2$~eV (95$\%$ confidence level, CL) \cite{Tanabashi:2018oca}. In this Letter we report on the first neutrino mass result from the Karlsruhe Tritium Neutrino experiment KATRIN \cite{Osipowicz:2001sq,KDR2004,Arenz:2018kma,Kleesiek:2018mel}, which is targeted to advance the sensitivity on \mnue\ by one order of magnitude down to 0.2~eV (90$\%$ CL) after~5 years.

\begin{figure*}[t!]
\begin{center}
\includegraphics[width=17cm]{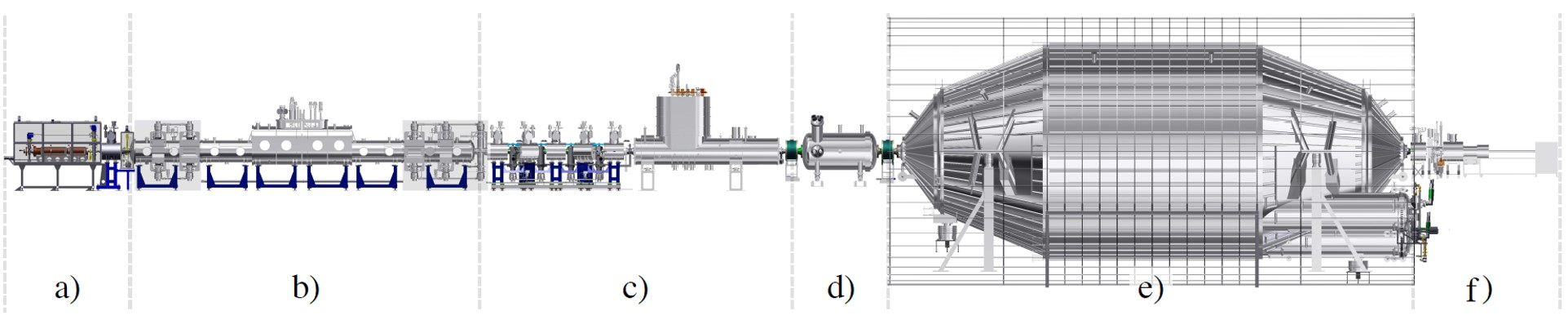}
\caption{\label{fig:katrin_setup} The major components of the KATRIN beam line consist of a) the Rear Section for diagnostics, b) the windowless gaseous tritium source WGTS, c) the pumping section with the DPS and CPS cryostats, and a tandem set-up of two MAC-E-filters: d) the smaller pre-spectrometer and e) the larger main spectrometer with its surrounding aircoil system. This system transmits only the highest-energy \bdecay\ electrons onto f) the solid-state detector where they are counted. 
}
\end{center}
\end{figure*}

{\em Experimental Setup}.-- KATRIN combines a windowless gaseous molecular tritium source (WGTS),
pioneered by the Los Alamos experiment  \cite{Robertson:1991vn}, with a spectrometer 
based on the principle of magnetic adiabatic collimation with electrostatic filtering (MAC-E-filter) \cite{Lobashev:1985mu, Picard1992}, developed at Mainz and Troitsk  \cite{Kraus2005,Aseev2011}. These techniques allow the investigation of the endpoint region of tritium \bdecay\ with very high energy resolution, large statistics and small systematics. KATRIN has been designed and built to refine this direct kinematic method to its ultimate precision level. To improve the sensitivity on \mnue\ by one order of magnitude calls for an increase in statistics and a reduction of systematic uncertainties by two orders of magnitude, as the observable in kinematic studies is the neutrino mass square, \mtwonue. 

Figure~\ref{fig:katrin_setup} gives an overview of the 70~m long experimental setup located at the Karlsruhe Institute of Technology (KIT). The source-related components in contact with tritium, the Rear Section RS (a), the source cryostat WGTS (b), as well as the differential (DPS) and cryogenic (CPS) pumping sections (c) are  integrated into the extensive infrastructure of Tritium Laboratory Karls\-ruhe to enable a closed cycle of tritium  \cite{Priester:2015bfa}. High-purity tritium gas from a pressure-controlled buffer vessel is continuously injected at 30~K into the WGTS at the midpoint of its 90~mm diameter, 10~m long stainless steel beamtube. The gas then diffuses  to both ends where it is pumped out by a series of turbomolecular pumps (TMPs) in the DPS, yielding the nominal column density $\rho d_{\mathrm{nom}}$ (5~$\cdot$~10$^{17}$ molecules cm$^{-2}$). In combination with the CPS, housing a large-capacity cryo\-trap operated at around 3~K, the flow rate of tritium into the following spectrometer and detector section (Fig. \ref{fig:katrin_setup} d-f) downstream is negligible, well below the 14~orders of magnitude of flow reduction required to eliminate source-related background by neutral tritium gas \cite{Osipowicz:2001sq}. 

The source magnetic field (B$_\mathrm{WGTS}$ = 2.52~T) as well as other superconducting solenoids \cite{Arenz:2018jpa} adiabatically guide primary \bdecay\ electrons, secondary electrons, and ions to the spectrometers. A series of blocking and dipole electrodes eliminates ions by an $\vec{E}$\,$\times$\,$\vec{B}$ drift to the beamtube, so that they cannot generate background in the spectrometer section \cite{Arenz:2018kma}.

High-precision electron spectroscopy is achieved by the MAC-E-filter technique, where electrons of charge $q$ are guided by the magnetic field, collimated by its gradient and filtered by an electrostatic barrier, the retarding potential energy $qU$. The resulting high-pass filter transmits only electrons with enough energy to overcome the barrier $qU$ and allows the scanning of the tritium $\upbeta$-decay spectrum in an integral mode.

The tandem configuration of MAC-E-filters performs a two-step filter process: first, the smaller pre-spectrometer is operated at fixed high voltage (HV) of $-10.4$~kV in this work to act as a pre-filter to reject electrons that carry no information on \mnue . In a second step, a variable $qU$ is applied to the main spectrometer for precision filtering of \bdecay\ electrons close to \ezero. Its huge size guarantees fully adiabatic motion to the central ``analyzing plane'', where the minimum magnetic field B$_\mathrm{min}$ and the maximum retarding energy $qU$ coincide for the filtering process to occur. Elevating the two spectrometers to a negative HV forms a strong Penning trap which can give rise to background \cite{Frankle:2014zja,Beck:2009ki}. This is avoided by operating both at an ultra-high vacuum (UHV) regime of 10$^{-11}$ mbar using non-evaporable getter (NEG) pumps and TMPs \cite{Arenz:2016mrh}.

A defining property of a MAC-E-filter is $\Delta E/E$, the filter width at energy $E$, which is given by the ratio $B_\mathrm{min}/B_\mathrm{max}$ of the minimum to maximum magnetic field in non-relativistic approximation. The present ratio (0.63~mT/4.24~T) is equivalent to $\Delta E = 2.8$~eV at \ezero. This value constrains the size V$_{\mathrm{ft}}$ of the flux-tube around $B_\mathrm{min}$ and, consequently, the overall background rate, which is proportional to V$_{\mathrm{ft}}$ to first order. A large aircoil system of 12.6 m dia\-meter \cite{Erhard:2017htg} is used to adjust $B_\mathrm{min}$ and  V$_{\mathrm{ft}}$. After the potential of the spectrometer vessel is elevated,  an offset of up to $-200$~V can be applied to the wire electrode system mounted on the inner surface of the vessel to define $qU$.

Electrons transmitted through the spectrometers are finally counted in a radially and azimuthally segmented monolithic silicon detector array with 148 pixels \cite{Amsbaugh:2014uca} as function of $qU$. To optimize the signal-to-background ratio, 
transmitted electrons are post-accelerated by a potential of +10~kV before they impinge on the detector.

{\em Commissioning measurements}.-- Over the past years we have commissioned the entire setup by a series of dedicated long-term measurements \cite{Arenz:2018kma,Arenz:2018jpa,Priester:2015bfa,Arenz:2018ymp} which have demonstrated that all specifications \cite{KDR2004} are met, or even surpassed by up to one order of magnitude, except for the background rate \rbg . 

A major benchmark is to operate the source at $\rho d_{\mathrm{nom}}$ at a stability level of $10^{-3}$/\,h so that variations of the column density $\rho d$ can be neglected.  This calls for a stable gas injection rate via capillaries \cite{Priester:2015bfa} and a constant beam-tube temperature. For the latter a stability level of better than 10$^{-3}$/~h has been achieved by a two-phase beam-tube cooling system at 30~(100)~K using neon (argon) as cooling fluid \cite{ Grohmann:2013ifa}. In mid-2018, measurements at 1~\% DT concentration within a 99~\% D$_2$ carrier gas at $\rho d_{\mathrm{nom}}$ have verified the required level of source stability \cite{KATRIN_first_tritium}. This ``first tritium'' campaign has allowed us to collect the first integral electron spectra which agree well with  the model expectation. 

In this spectral comparison the response function $f(E-qU)$ \cite{Kleesiek:2018mel} plays a fundamental role (see Eq.~\ref{eq:two}). It describes the probability of transmission of an electron with initial energy E as function of its surplus energy $E-qU$. For an ensemble, it depends on the angular spread of electrons and the amount of neutral gas they pass in the source, where they can undergo inelastic scattering processes with total cross section $\sigma$ (3.64~$\cdot$ 10$^{-18}$ cm$^2$ at 18.57~keV, adopted from \cite{liu1987}).

\begin{figure}[t!]
\begin{center}
\includegraphics[width=0.45
\textwidth]{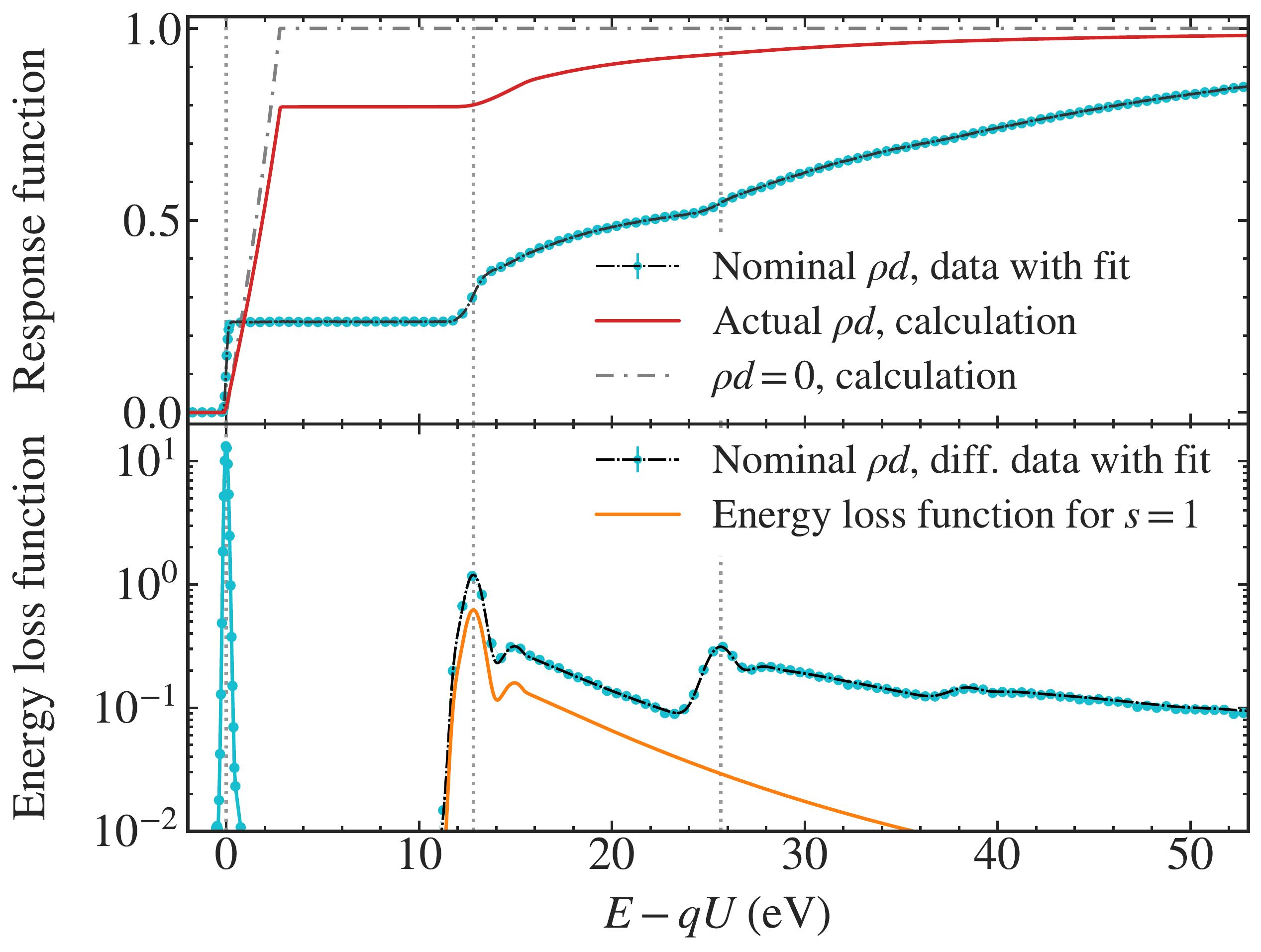}
\caption{\label{fig:response} 
(top) Measured and calculated response functions $f(E-qU)$ for electron surplus energies $E-qU$ at different $\rho d$ values of T$_2$. Measured $f(E-qU)$ for a narrow-angle photo-electron source close to $\rho d_{\mathrm{nom}}$ and fit (cyan); and calculated $f_{\mathrm{calc}}(E-qU)$ for isotropically emitted \bdecay\ electrons up to $\theta_{\mathrm{max}}$ at $\rho d_{\mathrm{exp}}$ (1.11 $\cdot$ 10$^{17}$ cm$^{-2}$), the set point of our scans (red line), and in the limit of vanishing $\rho d$ = 0 (grey, dash-dotted). (bottom) Differential distributions of energy losses $\delta$E from the MAC-E-ToF mode after a selection $35~\mu \mathrm{s} \leq ToF \leq 50~\mu$s at 
$\rho d \approx \rho d_{\mathrm{nom}}$
and fit (cyan). The ``no loss'' peak at $\delta E = E-qU = 0$ is followed by peaks with $s=2$ ($s=3$) scattering at twice (triple) the $\delta E$-value of $s=1$. The energy loss function \eloss\ for $s=1$ is obtained by deconvolution (orange). }
\end{center}
\end{figure}

We measure $f(E-qU)$ using monoenergetic electrons with a small angular spread produced in a dedicated photo-electron source (e-gun) \cite{Behrens:2017cmd} located at the RS. These electrons span a 50 eV wide range of surplus energies $E-qU$ and pass through the integral column density $\rho d$ of the source. This allows us to measure the characteristics of single ($s = 1$) and multiple  ($s = 2,3,...$) inelastic scattering. In Fig. \ref{fig:response} (top), we display the results for T$_2$ for the normal integrating MAC-E mode for $\rho d \approx \rho d_{\mathrm{nom}}$. The sharp rise with the filter width $\Delta  E$ to a plateau extending up to 11~eV results from ``no loss'' (energy loss $\delta E = 0$) e-gun electrons, which leave the source without scattering ($s = 0$) with a probability $\exp{(-\,\rho d\,\cdot\,\sigma)}$. At larger $E-qU$, $s$-fold scattering ($s = 1,2,3$) is visible. In Fig. \ref{fig:response} (bottom) the differential data from the 
MAC-E-ToF mode \cite{BONN1999256} are shown, where the electron time of flight (ToF) is recorded. This allows us to even better assess the $s$-fold inelastic scattering and to obtain the energy-loss function of electrons $\varepsilon$($\delta E$) by a deconvolution with the ``no loss'' peak at $\delta E = E-qU = 0$.

As the background rate \rbg\ exceeds its design goal of 0.01 counts per seconds (cps), we have studied the nature and origin of background processes so as to implement mitigation measures. Up to now, source-related backgrounds have not been observed, so that spectrometer-related processes \cite{Fraenkle:2017zpo} dominate \rbg , apart from a small detector-related contribution \cite{Amsbaugh:2014uca}. Electrons generated at the spectrometer surface by cosmic muons and environmental gamma rays are inhibited from entering the inner flux-tube by magnetic and electric barriers \cite{Arenz:2018aly,Altenmuller:2019xbg}. \rbg\ thus originates from  excited or unstable neutral atoms which can propagate freely in the UHV environment. Accordingly, \rbg\ is observed to have  an almost constant rate per unit volume in the flux-tube.

A significant part of \rbg\ is due to Rydberg atoms sputtered off the inner spectrometer surfaces by $^{206}$Pb-recoil ions following $\upalpha$-decays of $^{210}$Po. These processes follow the decay chain of the long-lived $^{222}$Rn progeny $^{210}$Pb, which was surface-implanted from ambient air (activity $\approx 1$ Bq/m$^2$) during the construction phase. A small fraction of these Rydberg states is ionized by black-body radiation when propagating over the magnetic flux-tube. The resulting sub-eV scale electrons are accelerated to $qU$ by the MAC-E-filter and form a Poisson component to \rbg . 

The other part stems from $\upalpha$-decays of single $^{219}$Rn atoms (t$_{1/2}$ =\,3.96 s) emanating from the NEG-pumps which
release a large number of electrons up to the keV-scale in the flux-tube, where they are stored due to its magnetic bottle characteristics. They subsequently produce secondaries until cooling off to energies of a few eV when they can escape 
and contribute to \rbg\ at $qU$. Owing to its origin from a small
number of $^{219}$Rn decays, this background includes a small non-Poissonian component \cite{Frankle:2011xy}. Liquid-nitrogen cooled copper baffles at the inlet of the NEG-pumps act as a countermeasure \cite{Goerhardt:2018wky}. Due to the formation of a thin layer of H$_2$O covering the baffle surface, the retention of $^{219}$Rn in this work is hampered such that \rbg\ retains a small non-Poissonian component. 

{\em Measurements of the tritium $\beta$-spectrum}.-- In the following we report on our first high-purity tritium campaign from April 10 to May 13, 2019 which demonstrates the functionality of all system components and of the extensive tritium infrastructure at large source activity ($2.45 \cdot 10^{10}$ Bq) and tritium throughput (4.9 g/day). As a result of radiochemical reactions of T$_2$ with the previously unexposed inner metal surface of the injection capillary we observe drifts in the source column density. To limit these drifts to a level of $\pm\ 2 \cdot 10^{-2}$ over our campaign, we keep the column density at an average value of $\rho d_{\mathrm{exp}} = 1.11~\cdot~10^{17}$~molecules~$\mathrm{cm}^{-2}$, which is about a factor of 5 smaller than $\rho d_{\mathrm{nom}}$.

At this setting, the smaller value of $\rho d_{\mathrm{exp}} \cdot \sigma$ (0.404) reduces the amount of inelastic scattering of electrons off neutral gas, see  Fig.~\ref{fig:response}. The relative fractions of the six hydrogen isotopologues injected into the source are continuously monitored by laser-Raman spectroscopy with $10^{-3}$ precision \cite{Schloesser2013a}. The average isotopic tritium purity $\varepsilon_{\mathrm{T}}$ (0.976) of our analyzed data sample is derived from the composition of the tritiated species T$_2$ (0.953), HT (0.035) and DT (0.011), with inactive species (D$_2$, HD and H$_2$) being present only in trace amounts.

Due to the large number of \bdecay s and ionization processes,  a cold magnetized plasma of electrons (meV to keV scale) and ions (meV scale) is formed which interacts with the neutral gas. The strong solenoidal field B$_\mathrm{WGTS}$ and the resulting large longitudinal conductance of the plasma allow the coupling of its potential to the surface of the Rear Wall (RW)  located at the RS and thus to control the starting energies of \bdecay\ electrons over the volume \cite{doi:10.13182/FST05-A1028}. Biasing the gold-plated RW disk with small areal variation of the work function to $-0.15$~V relative to the grounded beam tube gives a very good radial homogeneity of the source potential. This is verified during initial tritium scans with fits of  \ezero\ over detector pixel rings, which do not show a significant radial variation. 

Additional information on plasma effects is provided by comparing the line shape and position of quasi-monoenergetic conversion electrons (L$_3$-32) from \krm -runs   
in T$_2$ to \krm -runs without the carrier gas at 100~K \cite{Venos:2018tnw}. We do not identify sizeable shifts ($< 0.04$~eV) or broadening ($< 0.08$~eV) of lines so that the contribution of plasma effects at $\rho d_{\mathrm{exp}}$ to the systematic error budget in Table~\ref{tab:systematics} can be neglected.

\begin{figure}[t!]
\begin{center}
\includegraphics[width=0.45\textwidth]{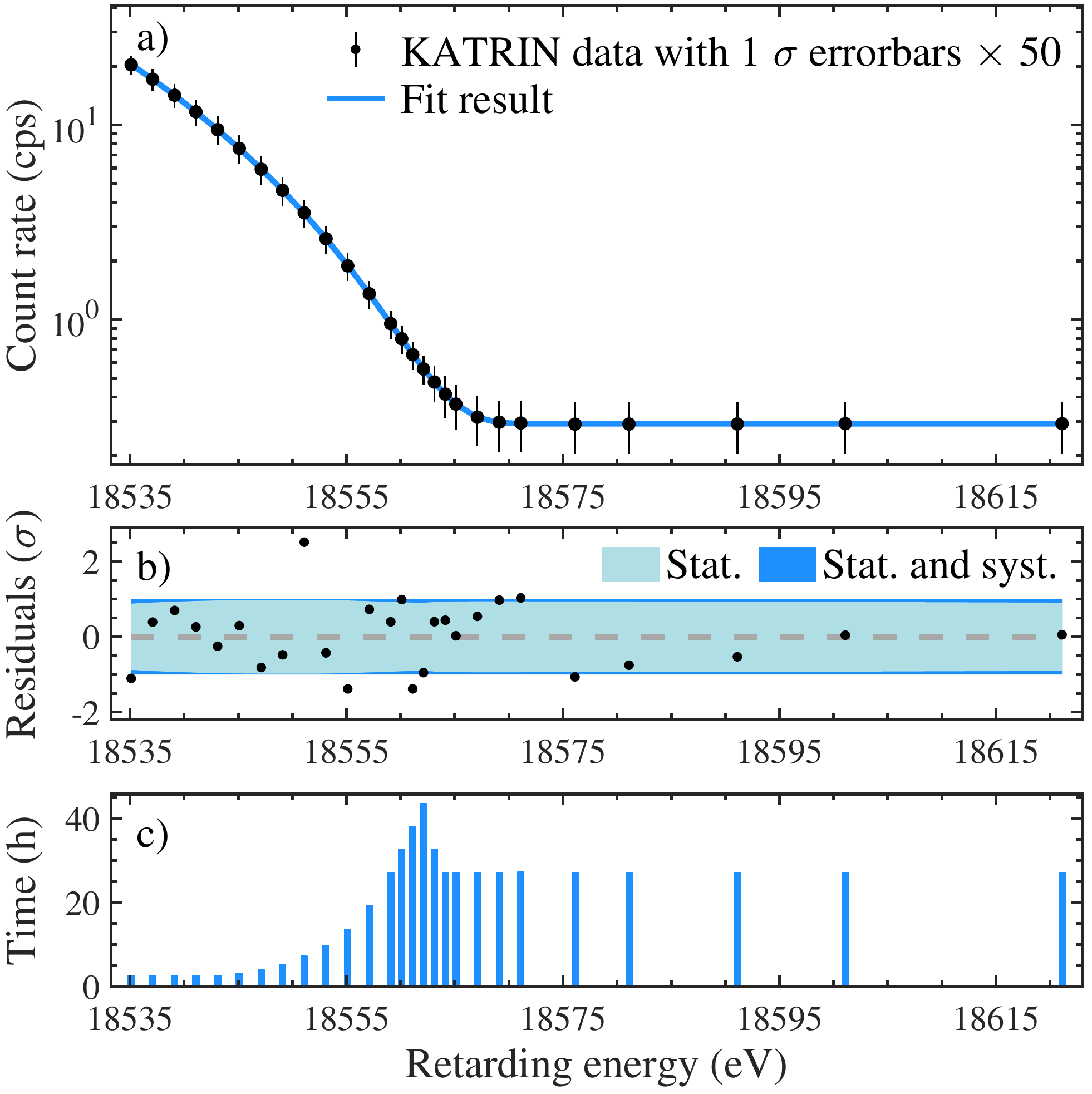}
\caption{\label{fig:beta_spectrum} a) Spectrum of electrons $R(\langle qU \rangle)$ over a 90~eV-wide interval from all 274 tritium scans and best-fit model 
$R_\mathrm{calc} (\langle qU \rangle)$ (line). The integral \bdecay\ spectrum extends up to \ezero\ on top of a flat background \rbg. Experimental data are stacked at the average value $\langle qU \rangle_l $ of each HV set point and are displayed with 1-$\sigma$ statistical uncertainties enlarged by a factor 50. b) Residuals of  $R(\langle qU \rangle)$ relative to the 1-$\sigma$ uncertainty band of the best fit model. c) Integral measurement time distribution of all 27 HV set points.
}
\end{center}
\end{figure}

The integral tritium \bdecay\ spectrum is scanned repeatedly in a range from $\lbrack \ezero - 90$ eV, $\ezero + 50$ eV$\rbrack$ by applying a set of non-equidistant HV settings to the inner electrode system. Each scan over this range takes a net time of about 2~h
and is performed in alternating upward and downward directions to compensate for any time-dependent drift of the system to first order. 
At each HV set point, the transmitted electrons are counted over time intervals varying from 17 to 576~s with typical values of $\sim 300$~s
for points close to \ezero. When setting a new HV value, we make use of a custom-made post-regulation system for
voltage stabilization and elimination of high-frequency noise. At the same time, a custom-made HV divider \cite{Thummler:2009rz} continuously monitors the retarding voltage with ppm precision.

For this work we analyze a scan range covering the region of 40~eV below \ezero\ (22 HV set points) and 50 eV above (5 HV set points). The non-uniform measuring time distribution in this interval is shown in Fig. \ref{fig:beta_spectrum} c). It maximizes the sensitivity for \mtwonue\
by focusing on the narrow region below \ezero, where the imprint of
the neutrino mass on the spectrum is most pronounced \cite{Kleesiek:2018mel}. Shorter time intervals with a set point 200~V  below \ezero ~are interspersed to monitor the source activity, in addition to other measures \cite{Babutzka:2012xd}.

{\em Data Analysis}.- For each tritium scan with its 27 HV set points, we apply quality cuts to relevant slow-control parameters to select a data set with stable run conditions. This results in 274 scans with an overall scanning time of 521.7~h. We also define a list of 117 detector pixels (out of 148), which excludes those pixels that are noisy or shadowed by  beamline instrumentation in the \belec\ path along the magnetic flux-tube. For the digitized, calibrated and pile-up-corrected detector spectra a broad region of interest (ROI) between 14 and 32~keV is defined. The ROI takes into account the detector energy resolution and its elevated potential (+10~kV) and allows us to include a large fraction of electrons backscattered at the detector in the narrow scan region close to \ezero\ \cite{Amsbaugh:2014uca}.

The long-term stability of the scanning process is verified by fits to single scans to extract their effective \bdecay\ endpoints. 
The 274 fit values show no time-dependent behavior and follow a Gaussian distribution ($\sigma$ = 0.25~eV) around a mean value. In view of this and the very good overall stability of the slow-control parameters for our data set, we merge the data of all 274~scans over all~117~pixels into one single 90-eV-wide spectrum, which is displayed in Fig. \ref{fig:beta_spectrum} a) in units of cps.

The underlying process corresponds to the ``stacking'' of events at the mean HV set points $\langle qU \rangle_l$ ($l = 1 - 27$). The small Gaussian spread (RMS = 34~mV) of the actual HV value $qU_{l,k}$ during a scan $k$ relative to $\langle qU \rangle_l$, the average of all scans, is a minor systematic effect which is accounted for in the analysis. The resulting stacked integral spectrum, 
R($\langle qU \rangle$), comprises $2.03 \cdot 10^6$ events, with $1.48\cdot 10^6$ \bdecay\ electrons below \ezero\ and a flat background ensemble of
$0.55\cdot 10^6$ events in the 90 eV scan interval. This high-statistics
data set allows us to show 1-$\sigma$ error bars enlarged by a factor of 50 in Fig. \ref{fig:beta_spectrum}. 

The experimental spectrum is well described by our detailed model of the KATRIN response to \bdecay\ electrons and background. It contains four free parameters: the signal amplitude $A_{\mathrm{s}}$, the effective \bdecay\ endpoint \ezero , the background rate $R_\mathrm{bg}$  and the neutrino mass 
square \mtwonue . We leave \ezero\ and $A_{\mathrm{s}}$ unconstrained, which is equivalent to a ``shape-only''
fit. The goodness-of-fit is illustrated in Fig.~\ref{fig:beta_spectrum} b) from the scatter of residuals around the error band of the model. 

The 4-parameter fit procedure over the averaged HV set points $\langle qU \rangle_l$ compares the experimental spectrum
R($\langle qU \rangle$) to the model $R_\mathrm{calc}(\langle qU \rangle)$. The latter is the convolution of the differential \belec\ spectrum \rbeta\ with the calculated response function $f_{\mathrm{calc}}(E - \langle qU \rangle)$, with an added energy-independent background rate $R_\mathrm{bg}$:

\begin{equation}  \label{eq:two}
R_\mathrm{calc} (\langle qU \rangle) = A_\mathrm{s}\,\cdot\,N_\mathrm{T}\int R_{\mathrm{\upbeta}}(E) \cdot  f_\mathrm{calc}(E - \langle qU \rangle)~ dE +  R_\mathrm{bg}~.
\end{equation}
Here, $N_\mathrm{T}$ denotes the number of tritium atoms in the source multiplied with the accepted solid angle of the setup $\Delta \Omega/4\pi = (1- \cos{ \theta_\mathrm{max}})/2$ and the detector efficiency ($\theta_{\mathrm{max}} = \arcsin{\sqrt{(B_\mathrm{WGTS}/B_\mathrm{max})}} = 50.4^{\circ}$).

The electron spectrum \rbeta\ from the superallowed \bdecay\ of molecular tritium is calculated using Fermi's Golden Rule: 
\begin{eqnarray}
   R_\upbeta(E) 
   = & & \frac{G_\mathrm{F}^2 \cdot \cos^2\Theta_\mathrm{C}}{2 \pi^3} \cdot \mtwohad \cdot F(E ,Z') \\ \nonumber
             &\cdot &  (E + \me ) \cdot \sqrt{(E + \me)^2 - \me^2} \\ \nonumber
    &\cdot & \sum_{\rm j} \zeta_j \cdot \varepsilon_j \cdot \sqrt{\varepsilon_{j}^2 - \mtwonue} \cdot \Theta(\varepsilon_{j}  - \mnue) ~ ,
\end{eqnarray}
with the square of the energy-independent nuclear matrix element $\mtwohad$, the neutrino energy $\varepsilon_j = \ezero - E - V_j$, the Fermi constant $G_\mathrm{F}$, the Cabibbo angle $\Theta_\mathrm{C}$, the electron mass $m_\mathrm{e}$, and the Fermi function $F(E,Z'=2)$. In addition, our calculations incorporate radiative corrections (for details see \cite{Kleesiek:2018mel,Drexlin2013}) and we account for thermal Doppler broadening at 30~K. 

When calculating \rbeta\ we sum over a final-state distribution (FSD) which is given by the probabilities $\zeta_j$ with which the daughter ion $^3$HeT$^+$ is left in
a molecular (i.\,e.\ a rotational, vibrational, and electronic) state with excitation energy $V_j$. For this analysis we first confirm the most recent theoretical FSD calculations \cite{Saenz2000,Doss2006} using new codes for solving the electronic
and rovibrational problems within the Born-Oppenheimer approximation. We then refine the FSD by adopting a more efficient treatment of
the rovibrational part and an update of other kinematics-related quantities, such as molecular masses, as well as recoil parameters (momenta and kinetic energy shifts). Most importantly, we treat all isotopologues (T$_2$, HT and DT) in a consistent way with initial angular momenta distributions $J_\kappa$ ($\kappa = 0,...,3$) at 30~K for the electronic bound states $n = 1,...,6$. The FSD includes higher excitation energies up to the continuum based on \cite{Saenz2000}, but their contribution to our analysis interval $\lbrack \ezero - 40$ eV$\rbrack$ is at an overall level of 10$^{-4}$ only. 
Accordingly, the FSD uncertainties in our narrow analysis interval of 40 eV below \ezero\ only contribute at the level of
0.02 eV$^2$ to the total systematics budget on \mtwonue\ (see Table \ref{tab:systematics}). 

\begin{table}[b!]
	\caption{1-$\sigma$ systematic uncertainties ($\sigma_{\mathrm{syst}}$) for \mtwonue\ in eV$^2$, averaged over positive and negative errors, using the method of MC propagation.}	
	\label{tab:systematics}
	\centering
	\begin{tabular}{p{4.3cm}p{2.4cm}p{1.5cm}}
	\hline
	Effect         & relative uncertainty &   $\sigma$(\mtwonue)\ ~~~~in eV$^2$ \\
	\hline\hline
	\textbf{Source properties} & &  \\
	$\rho d \cdot \sigma$ &  0.85\%  & 0.05\\
    energy loss $\varepsilon$($\delta$E) & $\mathcal{O}(1\%) $ & negligible  \\
    \hline
    \textbf{Beamline} & & 0.05 \\
    B$_\mathrm{WGTS}$ & 2.5 \% & \\  
    B$_\mathrm{min}$ & 1 \% & \\  
    B$_\mathrm{max}$ & 0.2 \% & \\     \hline
    \textbf{Final state distribution}  & $\mathcal{O}(1\%) $ & 0.02   \\
    \hline
    \textbf{Fluctuations in scan $k$} & & 0.05 \\
    HV stacking & 2 ppm & \\
	$\rho d$ variation & 0.8\% &  \\
    isotopologue fractions & 0.2\% & \\
    \hline
    \textbf{Background} & &  \\
    background slope & 1.7\%/keV & 0.07  \\
    non-Poisson background & 6.4\% & 0.30    \\
	\hline
	\textbf{Total syst. uncertainty} & & \textbf{0.32} \\
	\hline
	\end{tabular}
\end{table}

The response function $f_\mathrm{calc}(E-qU)$ used in the analysis is shown as the red curve in Fig. \ref{fig:response} (top). It corresponds to \bdecay\ electrons born with energies close to \ezero\ and emitted isotropically up to
$\theta_{\mathrm{max}}$ in the source gas. Compared to the e-gun beam, they possess a different distribution of energy losses due to their broader range of pitch angles $\theta$ and the varying amount of source gas ($\rho d$) they traverse. These processes are studied on the basis
of gas dynamical simulations \cite{Kuckert:2018kao} which yield an approximately triangular-shaped longitudinal source profile.

After modeling the energy loss of \bdecay\ electrons through the source by
making use of \rhods\ and $\varepsilon$($\delta E$), their subsequent propagation is tracked by the \textsc{Kassiopeia} simulation software \cite{Furse:2016fch}. It incorporates a detailed beamline model which takes account of the small radial inhomogeneities of $B_{\mathrm{min}}$ and $qU$ at the analyzing plane. The full model provides the detailed shape of $\Delta E$ and the distribution of electron pitch angles up to $\theta_{\mathrm{max}}$ from the parameters of the magnetic field triplet ($B_\mathrm{WGTS}$, $B_\mathrm{min}$, $B_\mathrm{max}$).

The energy-independent part of $R_\mathrm{calc}(\langle qU \rangle)$, \rbg, comes from a fit of the spectrum R($\langle qU \rangle$) over our 90 eV scan range. The fit value $R_\mathrm{bg} = (0.293 \pm 0.001)$~cps is largely constrained by the 5~HV set points above \ezero\ and agrees with data from independent  background runs.

The resulting model, $R_{\mathrm{calc}}(\langle qU \rangle)$, is then fitted to $R(\langle qU \rangle)$. To ensure that this proceeds without bias we employ a two-fold ``blinding''
scheme. The first blinding step leaves the data untouched, but a modification is applied during the building of the model $R_\mathrm{calc}(\langle qU \rangle)$. The FSD part describing rovibrational excitations of the electronic ground state is replaced with a Gaussian distribution with parameters not accessible to the analysis at first. As a result, fits with the blinded FSD do not reveal the unbiased value of \mtwonue . The ``true'' FSD is revealed only at the last step
(``unblinding'') after having fixed all model inputs and systematic uncertainties.

The second measure to mitigate biasing is to perform the full analysis, including parameter fitting, using Monte Carlo-based (MC) data sets first, before turning to the experimental data. For each experimental scan $k$ we generate a 
``MC twin'', $R_\mathrm{calc}$($\langle qU \rangle$)$_k$, from its averaged slow-control parameters to procure $R_{\upbeta}(E)_{k}$, $f_\mathrm{calc}(E - \langle qU \rangle)_k$ and $R_{\mathrm{bg},k}$. Analysis of ``MC twins'' allows us to 
verify the accuracy of our parameter inference by recovering the correct input MC-values for \mtwonue . This approach is also used to assess statistical ($\sigma_{\mathrm{stat}}$) and systematic ($\sigma_{\mathrm{syst}}$) uncertainties and to compute our expected sensitivity.

In the following we report on the results of two independent analyses with different strategies to propagate systematic uncertainties: 
 the ``Covariance Matrix'' and the ``MC propagation'' approaches. 

\begin{figure}[b!]
\begin{center}
\includegraphics[width=0.45\textwidth]{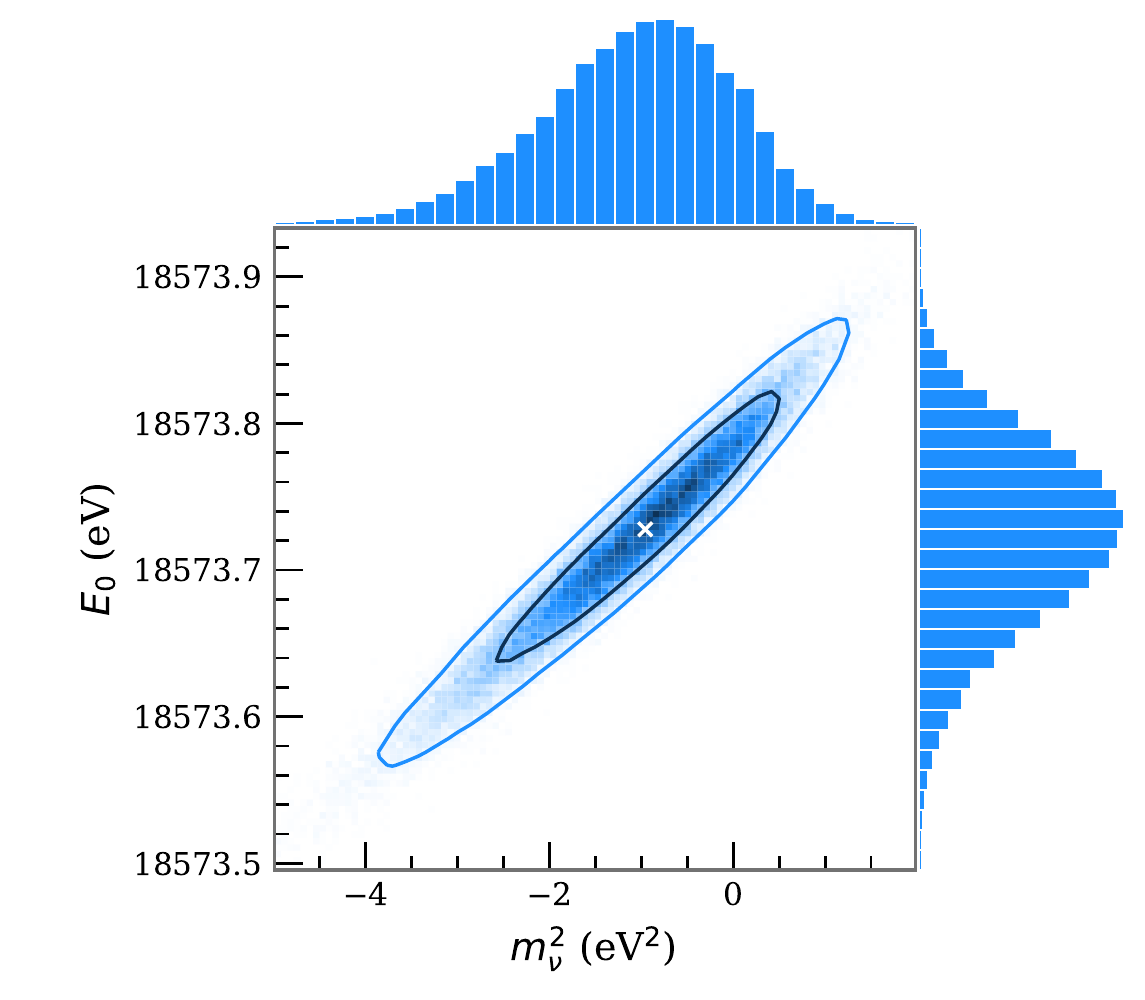}
\caption{\label{fig:correlation} Scatter plot of fit values for the mass square \mtwonue\  and the effective $\beta$-decay endpoint E$_0$ together with 1-$\sigma$ (black) and 2-$\sigma$ (blue) error contours around the best fit point (cross). It follows from a large set of 
pseudo-experiments emulating our experimental data set and its statistical and systematical uncertainties. 
}
\end{center}
\end{figure} 

In the covariance method we fit the experimental spectrum $R(\langle qU \rangle)$ with the model $R_\mathrm{calc} (\langle qU \rangle)$ by minimizing the standard $\chi^2$-estimator. To propagate the systematic uncertainties, a covariance matrix is computed after performing $O(10^4)$ simulations of $R_\mathrm{calc} (\langle qU \rangle)$, while varying the relevant parameters for each calculation according to the likelihood given by their uncertainties ~\cite{Barlow:213033,DAgostini:1993arp,KATRIN_first_tritium}. The resulting systematic uncertainties agree with the values shown in Table~\ref{tab:systematics}, which is based on the second approach. The sum of all matrices encodes the total uncertainties of $R_\mathrm{calc} (\langle qU \rangle)$ and their HV set point dependent correlations. The $\chi^2$-estimator is then minimized to determine the 4 best-fit parameters, and the shape of $\chi^2$-function is used to infer the uncertainties. The results of this fit are displayed in Fig.~\ref{fig:beta_spectrum}. We obtain a goodness-of-fit of $\chi^2 = 21.4$ for 23 d.o.f., corresponding to a p-value of 0.56.

The MC-propagation approach is a hybrid Bayesian-frequentist method, adapted from Refs.~\cite{Cowan:2010js,Cousins:1991qz,Harris:2014}. We fit the experimental spectrum $R(\langle qU \rangle)$ with the model $R_\mathrm{calc} (\langle qU \rangle)$ by minimizing the negative Poisson-likelihood function. The goodness-of-fit of $-2 \ln \mathcal{L}=23.3$ for 23 d.o.f. corresponds to a p-value of 0.44. To propagate the systematic uncertainties, we repeat the fit 10$^5$ times, while varying the relevant parameters in each fit according to their uncertainties given in column 2 of  Table~\ref{tab:systematics}.

We report the 1-$\sigma$ width of the  fit-parameters as their systematic uncertainty in the third column of Table \ref{tab:systematics}. In order to simultaneously treat statistical and all systematic uncertainties, each of the 10$^5$ fits is performed on a statistically fluctuated MC-copy of the true data set, leading to the distributions of \mtwonue\ and $E_0$ shown in Figure~\ref{fig:correlation}. The strong
correlation (0.97) between the two parameters is an expected feature in kinematic studies of \bdecay\ \cite{Drexlin2013,Otten2008}. The final-best fit is given by the mode of the fit-parameter distributions and the 1-$\sigma$ total error is determined by integrating the distributions up to 16\% from either side.

{\em Results}.-
The two independent methods agree to within a few percent of the total uncertainty. As best fit value for the neutrino mass we find \mtwonue\ = $(-1.0~ ^{+~ 0.9}_ {-~ 1.1})$ eV$^2$.  This best fit result corresponds to a 1-$\sigma$ statistical fluctuation to negative values of \mtwonue\ possessing a p-value of 0.16. 

The total uncertainty budget of \mtwonue\ is largely dominated by \sstat\ (0.97 eV$^2$) as compared to \ssyst\ (0.32 eV$^2$). As displayed in Table~\ref{tab:systematics}, the dominant contributions to \ssyst\ are found to be  the non-Poissonian background from radon and the uncertainty on the background slope, which is constrained from the wide-energy integral scans of the earlier ``first tritium'' data
\cite{KATRIN_first_tritium}. Uncertainties of the column density, energy-loss function, final-state distribution, and magnetic fields play a minor role in the budget of \ssyst. Likewise, the uncertainties induced via fluctuations of $\varepsilon_{\mathrm{T}}$ and HV parameters during a scan are negligibly small compared to \sstat.
The statistical (systematic) uncertainty of our first result on \mtwonue\ is smaller by a factor of 2 (6) compared to the final results of Troitsk and Mainz \cite{Aseev2011, Kraus2005}.

The methods of Lokhov and Tkachov (LT) \cite{Lokhov:2014zna} and of Feldman and Cousins (FC) \cite{Feldman:1997qc} are then used to calculate the upper limit on \mnue. Both procedures avoid 
empty confidence intervals for non-physical negative best-fit estimates of \mtwonue . For this first result we follow the LT method. For a statistical fluctuation into the non-physical region the method returns a confidence belt that coincides with the experimental sensitivity and avoids a shrinking upper limit for more negative values of \mtwonue . Using the LT construction we derive an upper limit of 
$\mnue < 1.1$~eV (90$\%$ CL) as the central result of this work. By construction it is identical to the expected sensitivity. For completeness we also note the FC upper limits $\mnue < 0.8~(0.9)$~eV at 90~(95)$\%$ CL. 

For the effective endpoint, our two analysis methods both obtain the best-fit value \ezero\ = ($18573.7~\pm~0.1$) eV 
(see Fig.~\ref{fig:correlation}). At this level of precision, a consistency check on the energy scale of KATRIN can be performed by comparing our experimental Q-value for molecular tritium with that based on measurements of the $^3$He-$^3$H atomic mass difference \cite{Myers2015}. 
Our result for the Q-value of $(18575.2 \pm 0.5)$~eV is obtained from our best-fit value for \ezero\ by adding the center-of-mass molecular recoil of T$_2$ (1.72~eV) \cite{Otten2008}, as well as the relative offset ($-0.2 \pm 0.5$ eV) of the source potential to the work function of the inner electrode. The calculated Q-value from the $^3$He-$^3$H atomic mass difference is $(18575.72 \pm 0.07)$~eV when accounting for the different binding energies and kinematic variables of atomic and molecular tritium \cite{Otten2008}. The consistency of both Q-values underlines the robustness of the energy scale in our scanning process of molecular tritium.

{\em Conclusion and outlook}.-- The reported upper limit $\mnue<1.1$~eV (90~\% CL) improves upon previous works \cite{Kraus2005,Aseev2011} by almost a factor of two after a measuring period of only four weeks
while operating at reduced column density. It is based on a purely 
kinematic method. As such it has implications for both particle physics and cosmology. For the former, it narrows down the allowed range of quasi-degenerate neutrino mass models by a direct method. For the latter, this model-independent limit can be used as laboratory-based input for studies of structure evolution in $\Lambda$CDM and other cosmological models. 

Our result shows the potential of KATRIN to probe \mnue\ by a direct kinematic method. After 1000 days of data taking at nominal column density and further reductions of systematics and \rbg , we will reach a sensitivity of 0.2 eV (90~\%\,CL) on \mnue , augmented by searches for physics beyond the SM, such as for sterile neutrino admixtures with masses from the eV to the keV scale.

\begin{acknowledgments}
We acknowledge the support of Helmholtz Association, Ministry for Education and Research BMBF (5A17PDA, 05A17PM3, 05A17PX3, 05A17VK2, and 05A17WO3), Helmholtz Alliance for Astroparticle Physics (HAP), Helmholtz Young Investigator Group (VH-NG-1055), 
Max Planck Research Group (MaxPlanck@TUM),
and Deutsche Forschungsgemeinschaft DFG (Research Training Groups GRK 1694 and GRK 2149, and Graduate School GSC 1085 - KSETA) in Germany; Ministry of Education, Youth and Sport (CANAM-LM2011019, LTT19005) in the Czech Republic; and the United States Department of Energy through grants DE-FG02-97ER41020, DE-FG02-94ER40818, DE-SC0004036, DE-FG02-97ER41033, DE-FG02-97ER41041, DE-AC02-05CH11231, DE-SC0011091, and DE-SC0019304, and the National Energy Research Scientific Computing Center.
\end{acknowledgments}  

\nocite{*}

\bibliographystyle{bst/apsrev4-2}
\bibliography{knm1}

\end{document}